\documentclass[draft,jgrga]{agutex}

  \usepackage[dvipdf]{graphicx}
  \usepackage[usenames]{color} 

  \setkeys{Gin}{draft=false}
\authorrunninghead{YERMOLAEV ET AL.}

\titlerunninghead{DYNAMICS SOLAR-WIND STREAMS. 2.}

\begin{document}

%
%

\title{Dynamics of large-scale solar-wind streams obtained by the double superposed epoch analysis. 
2. CIR.vs.Sheath and MC.vs.Ejecta comparisons 
}
%
%

%
%




\authors{Yu. I. Yermolaev, \altaffilmark{1}
I. G. Lodkina, \altaffilmark{1} 
N. S. Nikolaeva , \altaffilmark{1} 
M. Yu. Yermolaev \altaffilmark{1}
}

\altaffiltext{1}{Space Plasma Physics Department, Space Research Institute, 
Russian Academy of Sciences, Profsoyuznaya 84/32, Moscow 117997, Russia. 
(yermol@iki.rssi.ru)}






%


\begin{abstract} 

This work is a continuation of our previous paper 
\citep{Yermolaevetal2015} 
which describes the average temporal profiles of interplanetary plasma and 
field parameters in large-scale solar-wind (SW) streams: 
CIR, ICME (both MC and Ejecta) and Sheath as well as the interplanetary shock (IS). 
Like in the previous work we use data of OMNI database, 
our catalog of large-scale solar-wind phenomena during 1976--2000 
\citep{Yermolaevetal2009} 
and the double superposed epoch analysis (DSEA) method 
\citep{Yermolaevetal2010}: 
re-scaling the duration of interval for all types in such a manner that,
respectively, beginning and end for all intervals of selected type coincide. 
We present new detailed results of comparison of two pair phenomena: 
(1) both types of compression regions (CIR.vs.Sheath) and 
(2) both types of ICMEs (MC.vs.Ejecta). 
Obtained data allows us to suggest 
that the formation of all types of compression regions has 
the same physical mechanism irrespective of piston 
(High-Speed Stream (HSS) or ICME) type and 
differences are connected with geometry 
(angle between speed gradient in front of piston and satellite trajectory) 
and full jumps of speed in edges of compression regions. 
One of consequences of this hypothesis is the conclusion  
that one of the reasons of observed distinctions of parameters in MC and Ejecta 
can be fact that at measurements of Ejecta the satellite passes 
further from the nose area of ICME, than at measurements of MC. 
We also discuss the impact of Sheath in magnetospheric activity and 
its contribution in estimation of Sun's open magnetic flux.

\end{abstract}

%
%

%

\begin{article}

%
%

\section{Introduction} 

 

The disturbed types of the solar wind (SW) is 
one of the key links 
of space weather 
chain because 
only disturbed types of SW streams can contain the interplanetary magnetic field (IMF) component 
perpendicular to the ecliptic plane (in particular the southward IMF component) and 
be the main source of magnetospheric disturbances including the magnetic storms 
\citep{Russelletal1974,Burtonetal1975,Akasofu1981}.  
Such disturbed types are the following SW streams: 
interplanetary manifestation of coronal mass ejection (ICME) 
including magnetic cloud (MC) and Ejecta, 
Sheath - compression region before ICME, and 
corotating interaction region (CIR) - compression region 
before high-speed stream (HSS) of solar wind
(see reviews and recent papers by  
\cite{Gonzalezetal1999,HuttunenKoskinen2004,WimmerSchweingruberetal2006,ZurbuchenRichardson2006,Tsurutaniaetal2006,YermolaevYermolaev2006,YermolaevYermolaev2010,Yermolaevetal2007b,Yermolaevetal2012,Zhangetal2007,Jianetal2008,BorovskyDenton2010,ThatcherMuller2011,RichardsonCane2012,MitsakouMoussas2014,Hietalaetal2014,Cidetal2014,WuLepping2015,Katusetal2015,Kilpuaetal2015,Gopalswamyetal2015,Gopalswamyetal2016}
and references therein).
To understand geoeffectiveness of various types of SW streams, it is necessary to compare 
the characteristics of the streams inducing magnetospheric disturbances 
with the characteristics of all events of this type 
independently of possibility of disturbance generation.
In the present work we analyze full sets of various solar wind types for interval 1976-2000 
on the basis of OMNI data.

This paper is a continuation of our recent paper 
\citep{Yermolaevetal2015} 
in which we studied the average temporal behavior of plasma and field parameters 
of disturbed SW types by the double superposed epoch analysis (DSEA) method 
which allows one to investigate phenomena with different durations 
\citep{Yermolaevetal2010}.
To consider the influence 
of both the surrounding undisturbed SW types 
and the interaction of the disturbed SW types 
on the parameters, 
we separately analyzed the following sequences of the phenomena:
(1) SW/CIR/SW, 
(2) SW/IS/CIR/SW,
(3) SW/Ejecta/SW,
(4) SW/Sheath/Ejecta/SW,
(5) SW/IS/Sheath/Ejecta/SW, 
(6) SW/MC/SW,
(7) SW/Sheath/MC/SW,
(8) SW/IS/Sheath/MC/SW, 
where abbreviation IS means the interplanetary shock. 
This analysis allowed us to obtain several interesting results. 
In particular, we showed that 
the behavior of parameters in Sheath and in CIR are very similar 
both qualitatively and quantitatively.
Both the high-speed stream (HSS) and the fast ICME play a role of pistons 
which push the plasma located ahead them 
and result of compression does not depend on type of piston.
Besides, we obtained the clear evidence of ICME interaction 
with surrounding SW.  

In present work we continue our analysis and 
present new detailed results of comparison of two pair phenomena: 
(1) both types of compression region (CIR.vs.Sheath) and 
(2) both types of ICME (MC.vs.Ejecta). 
We also discussed the geoeffectiveness of compression regions 
and their contribution in the estimation of the Sun's open magnetic flux. 
The organization of the paper is as follows: Section 2 describes data and method. 
In Section 3, we present results on CIR.vs.Sheath and MC.vs.Ejecta comparisons. 
Section 4 discusses and summarizes the results.

\section{Methods} 

The data and methods used in this work are similar to those 
which have been used in the previous work 
\citep{Yermolaevetal2015}: 
the 1-h interplanetary plasma and magnetic field data of OMNI database 
(http://omniweb.gsfc.nasa.gov  
\citep{KingPapitashvili2004}), 
our catalog of large-scale solar-wind phenomena during 1976--2000 
(ftp://ftp.iki.rssi.ru/pub/omni/ 
\citep{Yermolaevetal2009}) 
and the double superposed epoch analysis (DSEA) method 
\citep{Yermolaevetal2010}: 
re-scaling the duration of the interval for all SW types in such a manner that, 
respectively, beginning and end for all intervals of selected type coincide. 
We selected only such events for which the SW type and 
the edges of an interval of the type could be defined on the basis of measurements 
(measurements of some parameters on this interval could be absent). 
Numbers of such events were 695 for Ejecta, 451 for CIR, 402 for Sheath, and 60 for MC. 

To characterize the value of parameters we use 
the terms "low", "high" and so forth. 
These terms are qualitative and are defined by comparison with the average value of 
the corresponding parameter in the undisturbed solar wind.
In order to estimate the existence of temporary change of parameter 
 in selected SW type, we defined the statistical significance of temporal trend 
as a linear dependence of parameter on time
\citep{BendatPiersol1971}. 
In all cases, when we write about the temporal change, 
there are linear dependences with probability not less 90\%.  

\section{Results}

\subsection{Comparison of CIR and Sheath phenomena}

Figures 1 and 2 present the average temporal profiles of 22 parameters 
of plasma and IMF for CIR (green lines), 
Sheath before Ejecta (blue lines) and Sheath before MC (red lines) 
obtained by the double superposed epoch analysis method. 
Data for phenomena with IS and without IS are 
shown by thick and thin lines, respectively. 

1. The bulk velocity $V$ (Fig.1e). 

The velocity $V$ increases for all types of solar wind: 
for phenomena without IS from $\sim$ 350--370 up to $\sim$430--470 km/s 
(the line for CIR has the highest slope) and 
for phenomena with IS from $\sim$ 420--450 up to $\sim$500--520 km/s 
(the slope for all phenomena is similar). 

2. The bulk velocity angles: longitude $\phi$ and latitude $\theta$ (Figs.1c, 1d). 
 
The angle $\phi$ increases from -3 -- -2 up to +2 degrees for CIR and 
from -2 up to +1 degrees for Sheath before Ejecta and 
is approximately constant for Sheath before MC. 
These dependences are similar both for phenomena with IS and without IS. 
The angle $\theta$ is approximately constant for all SW types and 
probably decreases from +3 down to -1 degrees only for Sheath before MC with IS.

3. The density $N$ (Fig.1b). 

The density decreases for all phenomena. 
For CIR and Sheath before Ejecta it falls rather sharply and monotonously, 
they with IS have higher value at the beginning of interval due to jump at IS 
but at the end of interval CIR and Sheath before Ejecta both with IS and without IS 
have similar density. 
The density for Sheath before MC with IS at the beginning of interval is close 
to one for CIR and Sheath before Ejecta with IS but it decreases slowly. 
The density for Sheath before MC without IS at the beginning of interval is close 
to one for CIR and Sheath before Ejecta without IS, 
then it increases in the middle of interval and 
decreases at the end of interval. 

4. The proton temperature $T_p$ (Fig.1a). 

For all SW types without IS the temperature grows from $5. 10^4$ K to $10. 10^4 K$, 
and for types of streams with IS it abruptly increases right after IS up to $10. 10^4 K$,
and further does not change on the interval.

5. The sound speed $V_s$ (Fig.1g).

The sound speed changes a little throughout the interval, 
but it is possible to note that it behaves similar to behavior of temperature 
in the corresponding SW types.

6. The Alfven speed $V_a$ (Fig.1f).

The Alfven speed poorly grows for all SW types, 
and for streams with IS value $V_a$ is a little higher. 

7. The thermal pressure $P_t$ (Fig.1h). 

The thermal pressure has characteristic features  of both temperature and density. 
At the beginning of interval the thermal pressure for phenomena with IS is 
one order of magnitude higher than phenomena without IS, 
then it in types with IS monotonic decreases and in type without IS increases  
in first half of interval and decrease a little in second half of interval.    

8. The ratio of measured and expected temperatures $T/T_{exp}$ (Fig.1i). 

The temperature ratio is larger 1 for all phenomena and 
it higher for phenomena with IS. 
The ratio for both Sheath types with IS increases at the IS front and 
then decreases throughout all interval and it has maxima inside interval 
for other SW types.  

9. The ratio of proton thermal and magnetic pressures ($\beta-$ parameter) 
(Fig.1j). 

For all phenomena the $\beta-$ parameter is almost constant 
(in the range 0.5--0.7) throughout interval and 
decreases only for Sheath before MC with IS. 

10. The magnitude of IMF $B$ (Fig.2b). 

For all phenomena without IS $B$ is lower than the for phenomena with IS and 
has maxima ($\sim$ 8 nT for CIR and Sheath before Ejecta and 
$\sim$ 12 for Sheath before MC) in the middle of interval. 
For CIR and Sheath before Ejecta with IS $B$ decreases from 12 down to 8 nT, 
and for Sheath before MC with IS it increases from 10 up to 14 nT.   

11. The x-, y- and z-components of IMF ($B_x$, $B_y$ and $B_z$) (Fig.2c,d,e).

For all phenomena the average values of $B_x$, $B_y$ and $B_z$ have small magnitude, 
are almost constant throughout all interval and change in the region of -2 -- +2 nT.  

12. The y-component of interplanetary electric field $E_y = V_x B_z$ (Fig.2f). 

For all phenomena the average value of $E_y$ has small magnitude, 
is almost constant throughout all interval and change in the region of -1 -- +1 mV/m.  

13. The sound Mach number $M_s$ (Fig.2a).
 
For all phenomena $M_s$ is almost constant in the region of 7--8 in first half interval 
and slightly increases in second part of interval. 

14. The Alfven March number $M_a$ (Fig.2g). 

For all phenomena $M_a$ slightly decreases from 10--12 down to 7--8 and 
it is almost constant $\sim$ 8 for Sheath before Ejecta with IS. 

15. The dynamic pressure $P_d$ (Fig.2h).
 
For all phenomena with IS $P_d$ is higher than phenomena without IS. 
For CIR and Sheath before Ejecta with IS $P_d$ decreases from 7 down to 4 dyn,
for they without IS it is almost constant $\sim 4$. 
For Sheath before MC without IS it increases from 3 up to 6 in first half of interval 
and does not change in second half of interval. 
For Sheath before MC with IS $P_d$ increases from 6 up to 8 dyn.

16. The magnetospheric $AE, K_p, D_{st}$ and $D^*_{st}$ indices (Fig.2i,j,k,l).
 
Behavior of all indices shows that magnetospheric activity increases throughout all interval. 
The activity is higher for phenomena with IS than phenomena without IS and 
activity for CIR is less than for all types of Sheath.

\subsection{Magnetic field in CIR and Sheath phenomena} 
The theoretical estimates of possible magnitudes of 
magnetic field in compression regions CIR and Sheath and 
their geoeffectiveness are discussed in the literature
(see papers by 
\cite{Gopalswamyetal2015,Gopalswamyetal2016} 
and references therein).
These estimates can be compared with results of measurements.  
Figure 3 presents the dependence of the average and 
maximal values of magnetic field $<B>$ and $B_{max}$ 
(left and right panels, respectively) on speed of corresponding piston:
on the speed of high-speed stream $V_{HSS}$ for CIR (green), 
on the velocity of Ejecta for Sheath before Ejecta $V_{Ej}$ (blue) and 
on the velocity of MC for Sheath before MC $V_{MC}$ (red). 
Straight lines are results of linear approximation of the 
corresponding data.
Thick lines and circles correspond to events with IS, 
thin lines and crosses to events without IS. 
Figure 4 has the same structure as figure 3, 
but unlike figure 3 it represents the average and maximum values 
of magnetic field 
related to the values of undisturbed magnetic field 
before compression region $<B>/B_{SW}$ and $B_{max}/B_{SW}$. 

Figure 3 shows that the average and maximal values of magnetic field 
$<B>$ and $B_{max}$ in all types of compression regions 
increase with increasing velocity of corresponding piston and 
can reach large values at high velocities of pistons. 
Though dependence of relative values $<B>/B_{SW}$ and $B_{max}/B_{SW}$
on speed of pistons is not observed, figure 4 shows 
that there are many cases 
when the average and maximal values of field in compression regions 
can exceed the field in undisturbed streams more than 4 times:
8 and 33 cases from 553 events 
for the average and maximal values of field, respectively. 

Two examples of such events of 18-21 December 1980 
(SW/IS/Sheath/MC/SW phenomena consequence) 
and 24-27 April 1979 (SW/IS/Sheath/Ejecta/SW consequence)
are presented in figures 5 and 6 
which have similar structure 
and show the following parameters: 
(panel a) the ratio of thermal and magnetic pressures ($\beta-$ parameter), 
the thermal pressure $P_t$, 
the ratio of measured and expected temperatures $T/T_{exp}$;
(b) the proton temperature $T_p$; 
(c) the solar wind velocity angles: longitude $\phi$ and latitude $\theta$;
(d) the z-component of IMF $B_z$ and 
the y-component of interplanetary electric field $E_y$;
(e) the measured and density-corrected $D_{st}$ and $D^*_{st}$ indexes; 
(f) the magnitude of IMF $B$, the dynamic pressure $P_d$;
(g) the y- and x-components of IMF ($B_y$ and $B_x$); 
(h) the sound and Alfvenic velocities $V_s$ and $V_a$; 
(i) the ion density $N$, the $K_p$ index increased by coefficient 10; 
(j) the solar-wind bulk velocity $V$, the $AE$ index. 
OMNI dataset has 7h gap for several plasma parameters after IS on 19 December, 1980 
but there is total set of IMF measurements. 

Both examples show that the magnitude of IMF $B$ has jump at IS 
from $\sim 5$ up to $\sim 12-15$ nT (factor $\sim 2-3$) 
and then $B$ continues to grow up to $\sim 35$ nT (total factor $\sim 7$). 
In first case the southward component of IMF $|B_z|$ increases up to $\sim 30$ nT
and the compression region Sheath before MC induced the strong magnetic storm 
with $D_{st} \sim -250$ nT. 
In second case the southward component $|B_z|$ has two short maxima $\sim 17$ nT 
and the compression region Sheath before Ejecta induced the multi-step magnetic storm 
with $D_{st} \sim -150$ nT. 
It is important to note that in this case 
the maximal magnitude of IMF $B$ in Sheath 
is significantly higher (factor $>2$) than in Ejecta. 

\subsection{Comparison of MC and Ejecta phenomena}

Figures 7 and 8 present the temporal profile of 22 SW plasma and IMF parameters 
for MC (with IS+Sheath, with Sheath and without IS+Sheath - 
thick, thin and dash red lines, respectively), 
and Ejecta (with IS+Sheath, with Sheath and without IS+Sheath - 
thick, thin and dash blue lines) obtained
by the double superposed epoch analysis. 

1. The bulk velocity $V$ (Fig.7e). 

The velocity $V$ decreases for all ICME types and 
it is higher for phenomena with IS 
than for phenomena without IS. 
The velocity differences between subtypes of MC is small and 
it for subtypes of Ejecta is large: 
the velocity difference between Ejecta and Ejecta with Sheath is $\sim 70$ km/s and 
the velocity difference between Ejecta and Ejecta with Sheath and IS is $\sim 100$ km/s 
and these differences are larger at the beginning of interval than at the end.  

2. The bulk velocity angles: longitude $\phi$ and latitude $\theta$ (Figs.7c, d).

The angle $\phi$ decreases from $\sim 2$ down to $\sim 0$ degrees for Ejecta 
with Sheath and IS, 
from $\sim 1$ down to $\sim 0$ degrees for Ejecta with Sheath and without IS 
and is almost 0 for Ejecta without Sheath. 
The angle $\theta$ is $\sim 1$ degrees for  all subtypes of Ejecta.
For MC the angles $\phi$ and $\theta$ have large spread 
(probably because of small statistics) and a tendency can not be estimated. 
 
3. The density $N$ (Fig.7b). 

The density for all types (except MC with Sheath and without IS) has minima  
in the middle of interval 
and minimal value $\sim 5$ cm$^{-3}$ is observed for Ejecta with Sheath and IS. 
Phenomena with IS have abrupt jumps in the beginning of interval 
(at the IS front) by a factor of $\sim 5$, 
other phenomena in the beginning of interval and 
all phenomena in the end of interval have maxima by a factor of  $\sim 2$. 
 
4. The proton temperature $T_p$ (Fig.7a). 

The temperature for all types decreases in the first half of interval and 
is almost constant in the second part of interval. 
Phenomena with IS have jumps in the beginning of interval (at the IS front). 
The temperature for all subtypes of Ejecta is higher than for similar subtypes of MC.

5. The sound speed $V_s$ (Fig.7g). 

The sound speed changes a little 
throughout the interval, 
but it is possible to note that it behaves similar to behavior 
of temperature in the corresponding SW types.

6. The Alfven speed $V_a$ (Fig.7f).
 
The Alfven speed is higher for all subtypes of MC then for Ejecta. 
It is almost constant for Ejecta and 
it is lowest for Ejecta without Sheath and IS and 
highest Ejecta with Sheath and IS ($\sim 70$ km/s). 
The Alfven speed has maximum $\sim 120$ km/s for MC with IS 
in the middle of interval.
  
7. The thermal pressure $P_t$ (Fig.7h). 

The thermal pressure is closed for phenomena Ejecta and MC, 
it increases in the ends of interval 
(there is a jump at the IS front for phenomena with IS). 
  
8. The ratio of measured and expected temperatures $T/T_{exp}$ (Fig.7i). 

The ratio of temperatures is closed for phenomena Ejecta and MC,  
$T/T_{exp} < 1$, and it increases in the ends of interval 
(there is a jump at the IS front for phenomena with IS). 

9. The ratio of proton thermal and magnetic pressures ($\beta-$parameter) 
(Fig.7j). 

The ratio is almost constant for all subtypes E and $\beta-$parameter $ = \sim 0.5$. 
The ratio has minimum $\sim 0.1$ for MC in the middle of interval. 

10. The magnitude of IMF $B$ (Fig.8b). 

In accordance with definition of Ejecta and MC, 
the magnitude of IMF is higher for MC (maximum $\sim 15$ nT for MC with IS) 
than for Ejecta (minimum $\sim 6$ nT for Ejecta without Sheath). 
For phenomena with IS there is jump at the front of IS and 
$B$ decreases throughout all interval. 
 
11. The x-, y- and z-components of IMF ($B_x$, $B_y$ and $B_z$) (Fig.8c,d,e). 

The average values of IMF components is closed to zero for all phenomena, 
but for MC with IS $B_z < 0$ in the beginning of interval and 
$B_z > 0$ in the end of interval.  

12. The y-component of interplanetary electric field $E_y$ (Fig.8f). 

The electric field variation is similar to behavior of IMF $B_z$ component 
for corresponding SW type.

13. The sound Mach number $M_s$ (Fig.8a). 

The sound Mach number decreases for all phenomena throughout all interval. 
Values $M_s$ is higher for MC than for Ejecta and 
higher for phenomena with IS than for phenomena without IS.

14. The Alfven March number $M_a$ (Fig.8g). 

The Alfven March number is low for all phenomena. 
Values $M_a$ is higher for Ejecta than for MC and 
lower for phenomena with IS than for phenomena without IS.

15. The dynamic pressure $P_d$ (Fig.8h).
 
The dynamic pressure is almost constant for all phenomena 
(phenomena with IS have a jump in the beginning of interval) and 
it is higher for MC than for Ejecta.  

16. The magnetospheric $AE, K_p, D_{st}$ and $D^*_{st}$ indices (Fig.8i,j,k,l). 

All indices show that magnetospheric activity decreases throughout all interval. 
The activity is higher for MC than for Ejecta and 
for phenomena with IS than phenomena without IS.    

\section{Discussion and Conclusions} 

As has been noted in our first article 
\citep{Yermolaevetal2015}, 
in general, our results on compression regions are close to earlier published data 
(e.g. 
\cite{Jianetal2008,BorovskyDenton2010,MitsakouMoussas2014} 
and references therein), 
but unlike the previous works we carried out additional selection of events 
by types of pistons and existence of IS. 
In the present work we compare several key parameters in more detail 
taking into account this selection. 
For the majority of parameters their behavior is close for different pistons, 
but depends on the existence of IS. 
For example, all types of compression regions have on average identical speed profiles 
both in value and in inclination, 
and the increase of speed of pistons on 100 km/s leads to formation of IS. 
For some parameters the main difference consists in their jump at the beginning of interval. 

Obtained data allows us to suggest that 
the formation of all types of compression regions has the same physical mechanism
irrespective of piston type. 
To check this hypothesis, it would be useful to compare not only full jumps of speed 
in compression regions (see figure 1e), 
but also speed gradients in these regions. 
It should be noted that the presented average temporary profiles of speed do not 
allow one to directly define the average gradients of speed 
for various types of compression regions 
as measurements are carried out under unknown angle relative to the speed gradient. 
If making the natural assumption that the gradient of speed is directed approximately 
on normals to the piston, 
CIR has the greatest angle between the gradient of speed and 
the direction of average SW speed,
and ICME - the smallest angle (in order that the satellite can cross ICME body, 
it has to be near a nose area of ICME). 
The average real (before the rescaling during DSEA processing) durations of 
compression regions considerably differ: 
for figures 1 and 2 we took  duration of CIR of 20 hours (close to real), 
and the average real durations of Sheath before Ejecta and Sheath before MC 
are shorter by a factor of $\sim1.5$ and $\sim2$, respectively 
(e.g. 
\cite{Yermolaevetal2010b}). 
This duration difference is qualitatively agree with the assumption 
that all types of compression regions have approximately identical sizes 
in the direction of speed gradient, 
and taking into account the identical total jumps of speed in compression regions 
(see figure 1e)
they have also identical gradients of speed. 
One of consequences of this hypothesis is the conclusion 
that one of the reasons of observed distinctions of parameters in MC and Ejecta 
can be fact that at measurements of Ejecta the satellite passes further 
from the nose area of ICME, 
than at measurements of MC, 
i.e. distinctions can be connected partially with conditions of observation, 
but not with physical distinctions between Ejecta and MC.
This result agrees with geometrical selection of ICME measurements described by 
\cite{Jianetal2006}.  

Increase of piston speed increases the magnitude of magnetic field and its components 
in the compression regions and, as a result, magnetospheric disturbances. 
As show the presented results the IMF magnitude can reach rather high value, and 
ratios of average and maximum values to level of undisturbed SW can be $>4$ 
in 1.5 and 6\% of cases, respectively. 
This percentage of events is close to percentage of strong magnetic storms 
\citep{Yermolaevetal2013}. 
At average dependences of B on piston speed (see figures 3 and 4) 
the SW speeds observed in experiments can lead to great values of IMF field. 
So at a speed in HSS more than 1000 km/s the magnetic field in CIR can increase to 50--70 nT,
and at a speed of MC $\sim 2000$ km/s (as, for example, in events of August of 1972 
\citep{VaisbergZastenker1976}) 
up to 100 nT.
It should be note that efficiency of magnetic storm generation is 
$\sim50$\% higher for Sheath and CIR than for ICME (MC and Ejecta)
\citep{Nikolaevaetal2013,Nikolaevaetal2015}, 
i.e. at identical southward components of interplanetary field 
the magnetic storms are generated $\sim1.5$ times more strongly by Sheath and CIR
than by ICME. 
Thus it is possible to conclude that in our opinion 
the role of compression regions (especially Sheath) in generation of storms 
is often underestimated. 

On the other hand, the wrong estimation of contribution of 
Sheath magnetic field in the IMF can lead to mistakes 
in studying of magnetic field of the Sun. 
Several recent works proposed that the Sun's open magnetic field flux 
consists of a time-independent (floor) component 
which may be observed during solar minimum and 
a time-varying component due to CMEs (see recent papers by 
\cite{WangSheeley2015,Cliver-vonSteiger2015} 
and references therein). 
Because the Sheaths originate from outside the CMEs in the Sun, 
the inclusion of  the Sheaths in the ICMEs 
(see, for example, catalog by 
\cite{RichardsonCane2010}) 
can result in significant overestimation of the contribution 
of the CME itself to the IMF and 
incorrect estimation of Sun's open magnetic flux on the basis of 
interplanetary measurements 
\citep{Yermolaevetal2009b}. 

It would be desirable to pay attention of the reader to the interesting fact: 
at the analysis of all events (without selection by their geoefficiency)
the magnetosphere activity for CIR/Sheath increases in interval 
but for MC/Ejecta it increases in the beginning of interval and then decreases.  

Figures 7 and 8 show that all ICME types have several traces of compression 
near interval edges (more strongly on the leading edge and 
more weakly on the training edge): 
(1) the increase of density $n$, 
(2) the  of temperatures $T_p$ and $T_p/T_{exp}$, 
(3) the increase of dynamic and thermal pressures $P_d$ and $P_k$, 
(4) the increase of $\beta-$parameter, and 
(5) the turn of speed angle $\phi$.  
These facts confirm the ICME expansion and its interaction 
with surrounding solar wind.  
As the average velocity of ICME is higher than that of surrounding solar wind, 
the compression on the leading edge is higher than on the training edge.




%
%
%
%
%
%
%

\begin{acknowledgments}
The authors are grateful for the opportunity to use the OMNI database. 
The OMNI data were obtained from GSFC/ SPDF OMNIWeb (http://omniweb.gsfc.nasa.gov). 
YY is grateful to SCOSTEP's "Variability of the Sun and Its Terrestrial Impact" 
(VarSITI) program for the support of his participation in the workshop 
"International Study for Earth-Affecting Solar Transients (ISEST)/MiniMax" in Mexico City, Mexico, October 26 -- 30, 2015. 
This work was supported by the 
Russian Foundation for Basic Research, project 
16--02--00125, and by Program 7 of Presidium of the Russian Academy of Sciences. 

\end{acknowledgments}

\end{article}


%
%

%
%
%
%
%


\begin{figure}
\noindent\includegraphics[width=14cm]{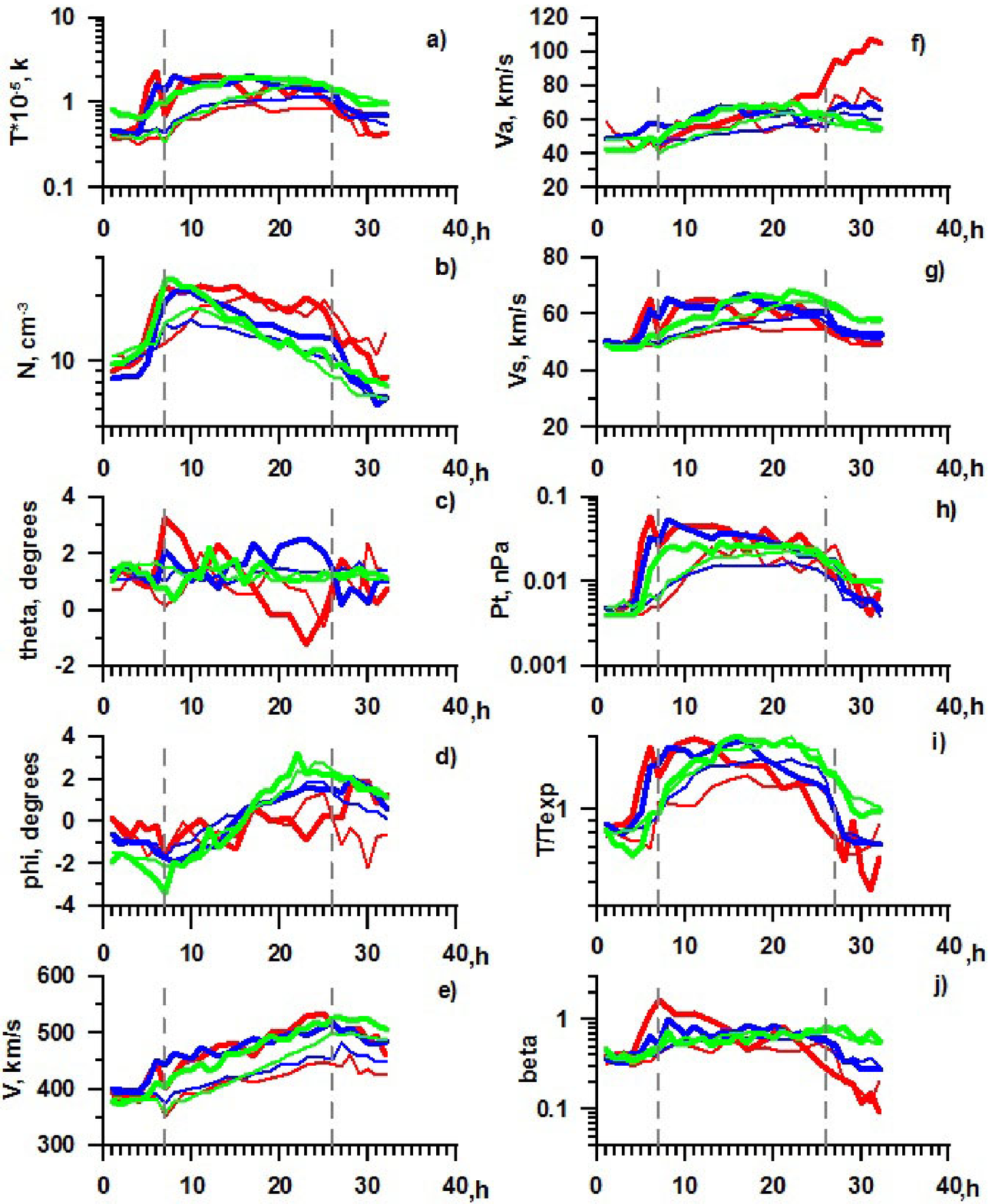}
\caption{The temporal profile of SW plasma and IMF parameters for CIR 
(with IS and without IS - bold and thin green lines, respectively) and 
Sheath before Ejecta (with IS and without IS - bold and thin blue lines) and 
Sheath before MC (with IS and without IS - bold and thin red lines) 
obtained by the double superposed epoch analysis}
\end{figure}


\begin{figure}
\noindent\includegraphics[width=14cm]{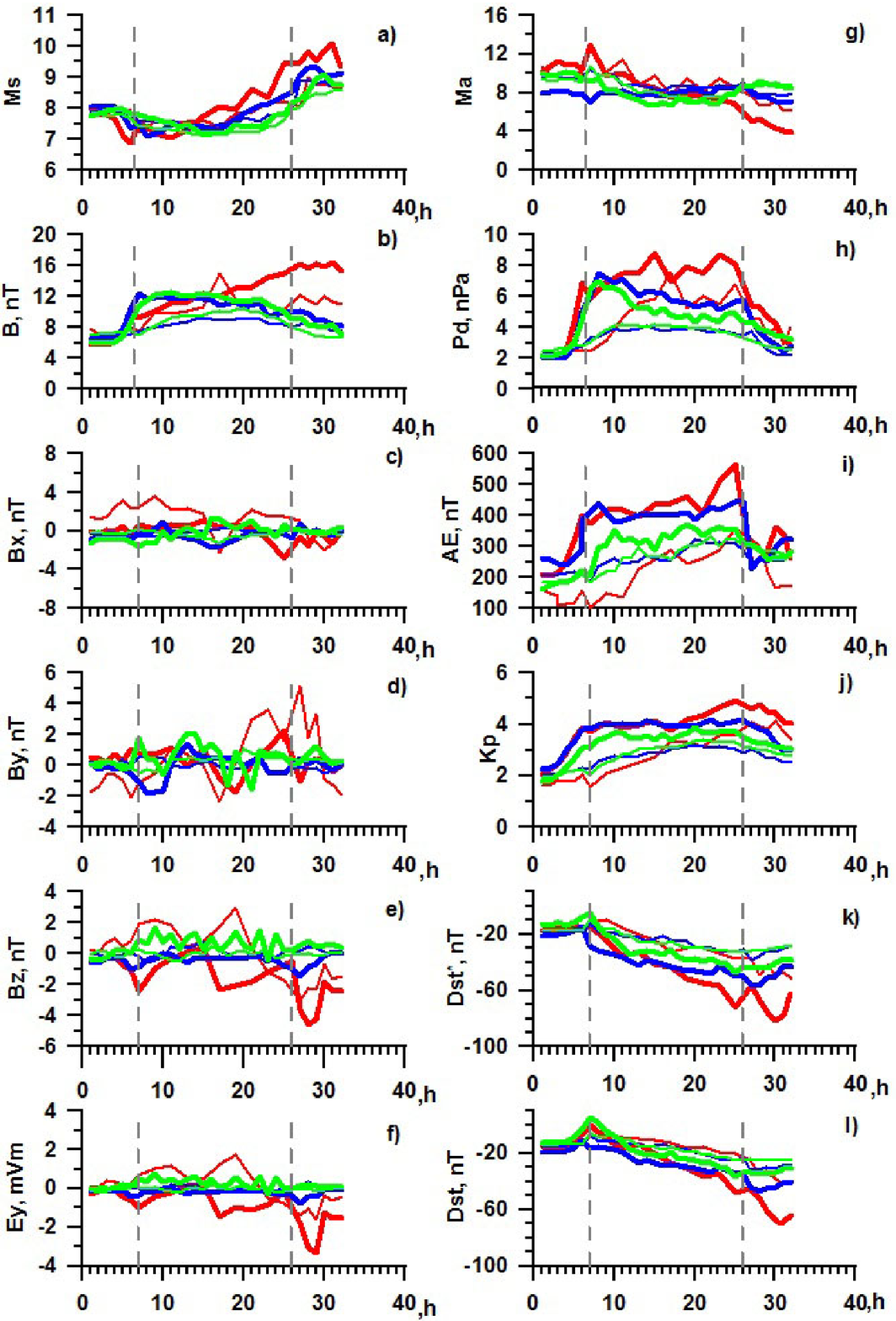}
\caption{Continuation of figure 1}
\end{figure}


\begin{figure}
\noindent\includegraphics[width=14cm]{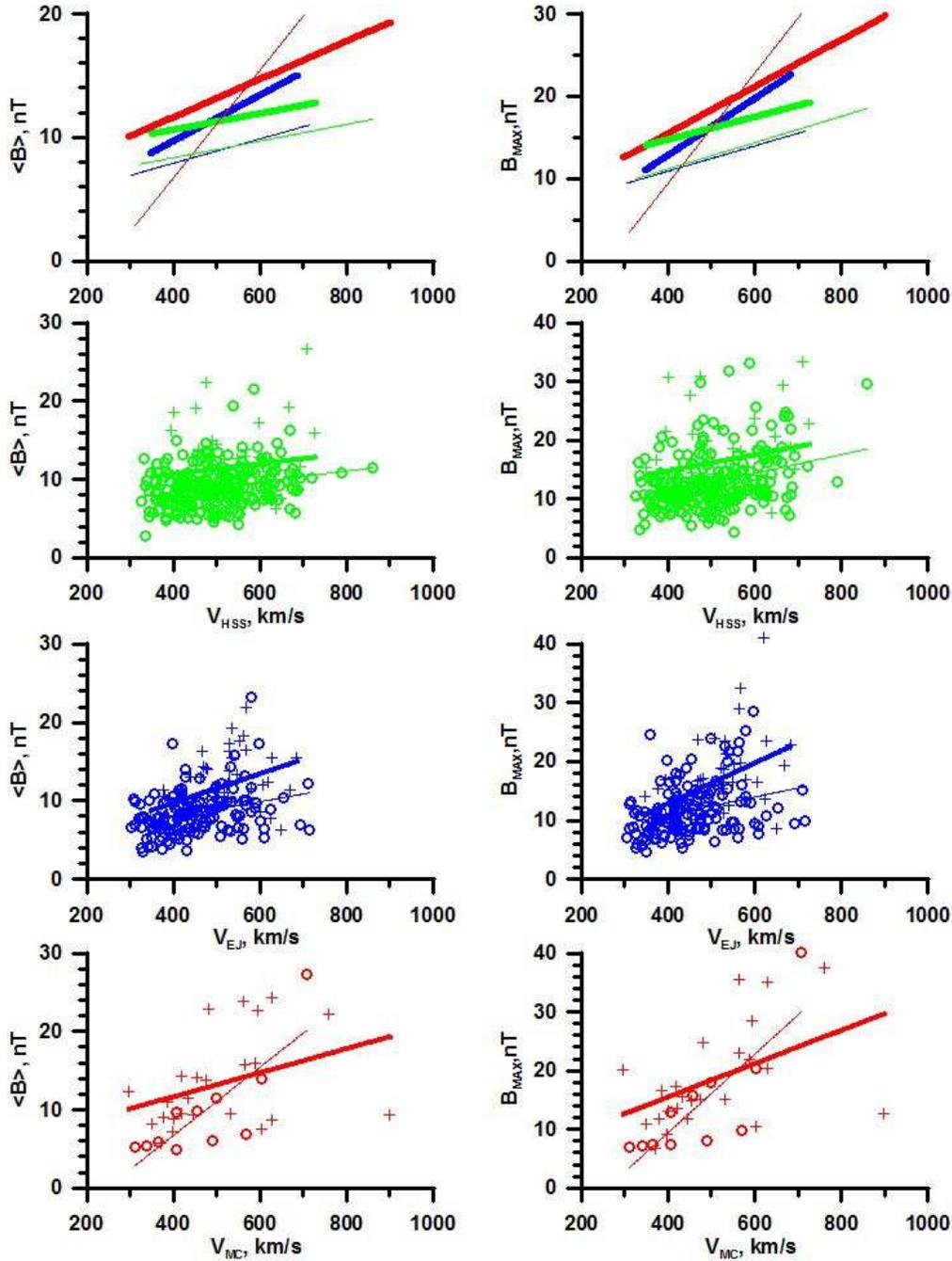}
\caption{Dependence of the average and maximal values of magnetic field $<B>$ and $B_{max}$ 
on the speed of piston for CIR (green), Sheath before Ejecta (blue) and Sheath before MC (red). 
Bold lines and circles for events with IS, thin lines and crosses for events without IS}
\end{figure}


\begin{figure}
\noindent\includegraphics[width=14cm]{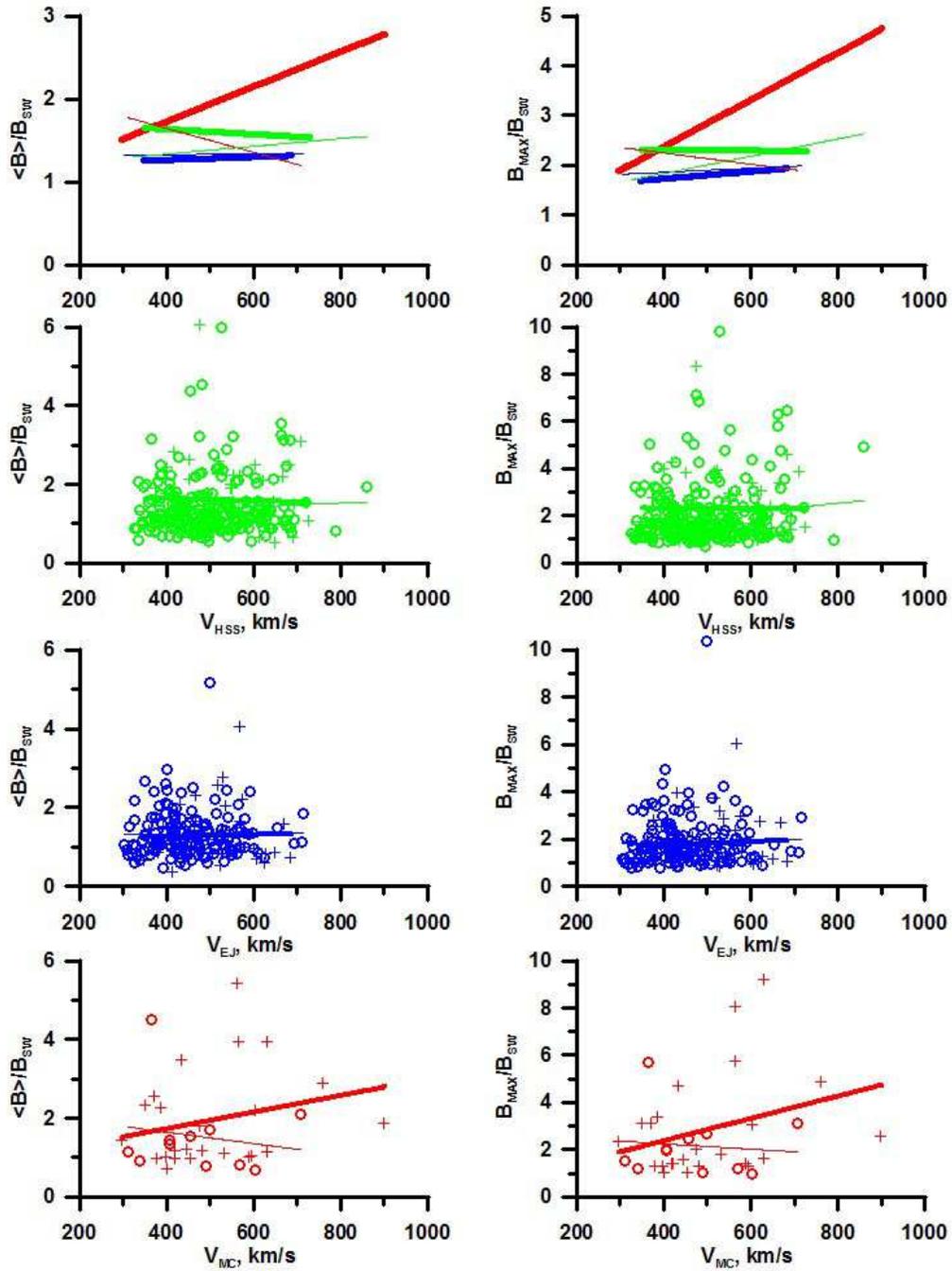}
\caption{The same dependence of normalized values $<B>/B_{SW}$ and $B_{max}/B_{SW}$}
\end{figure}

\begin{figure}
\noindent\includegraphics[width=14cm]{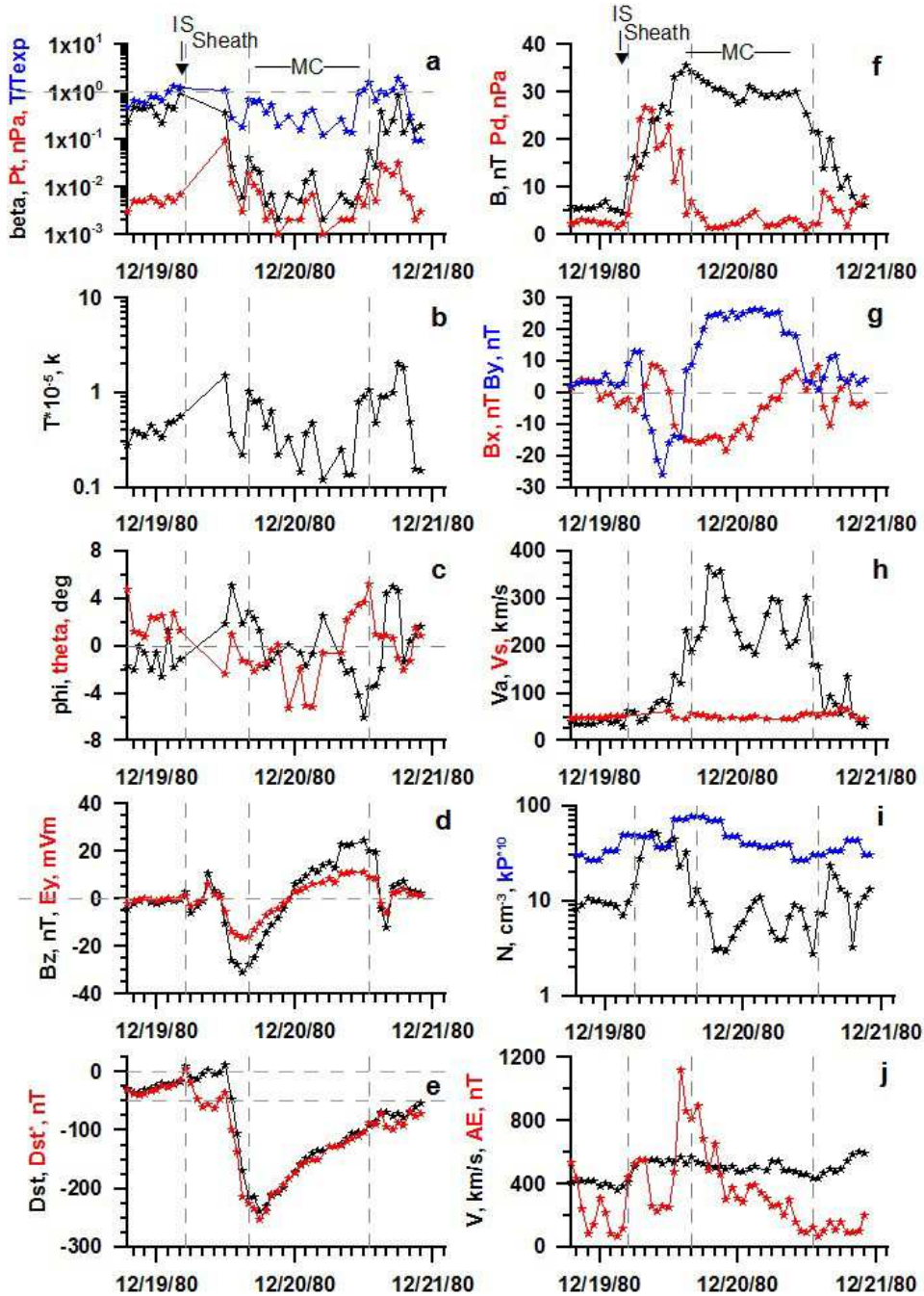}
\caption{The temporal profile of SW plasma and IMF parameters for consequence SW/IS/Sheath/MC/SW in period of 18-21 December 1980}
\end{figure}

\begin{figure}
\noindent\includegraphics[width=14cm]{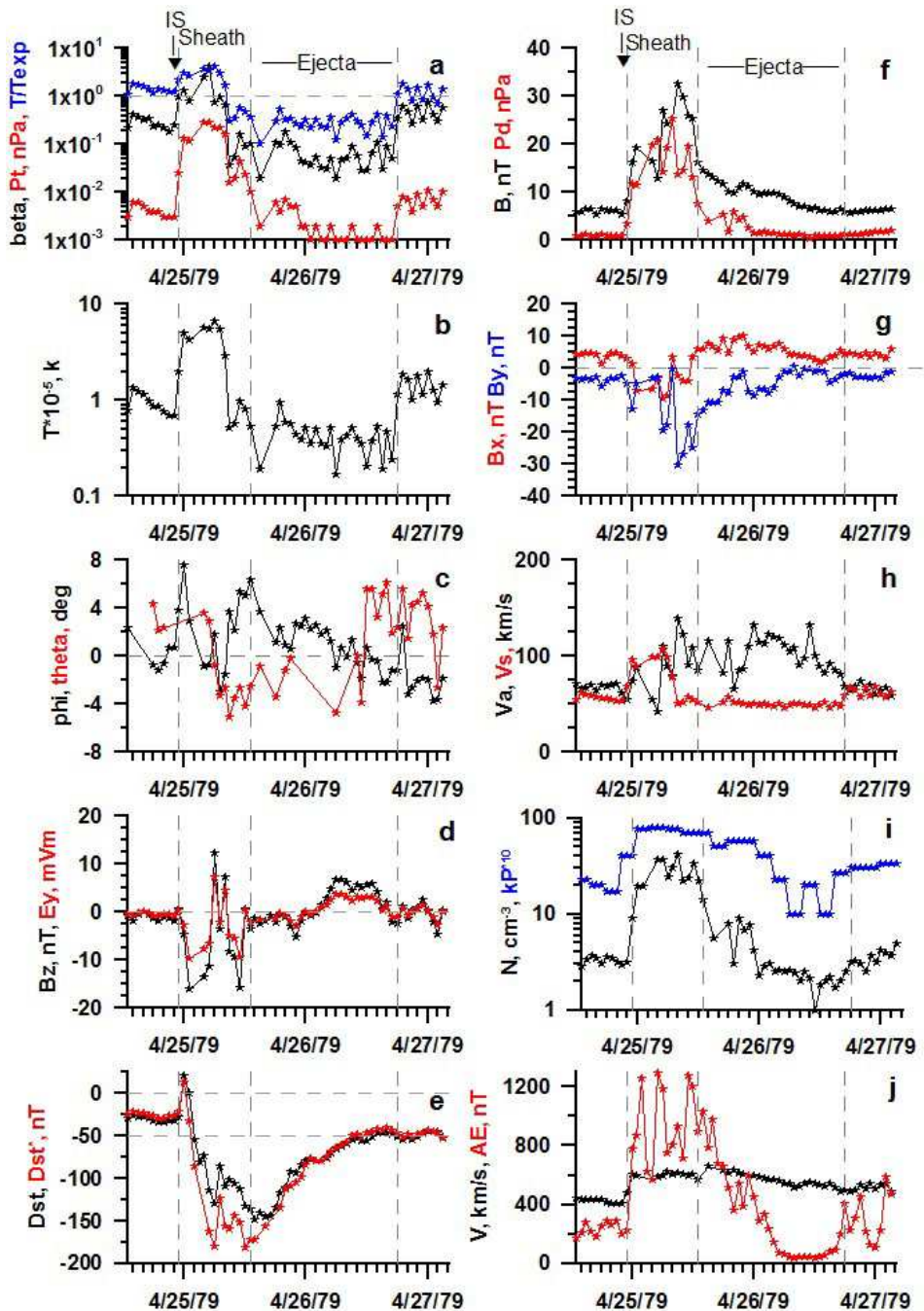}
\caption{The temporal profile of SW plasma and IMF parameters for SW/IS/Sheath/Ejecta/SW in period of 24-27 April 1979}
\end{figure}

\begin{figure}
\noindent\includegraphics[width=14cm]{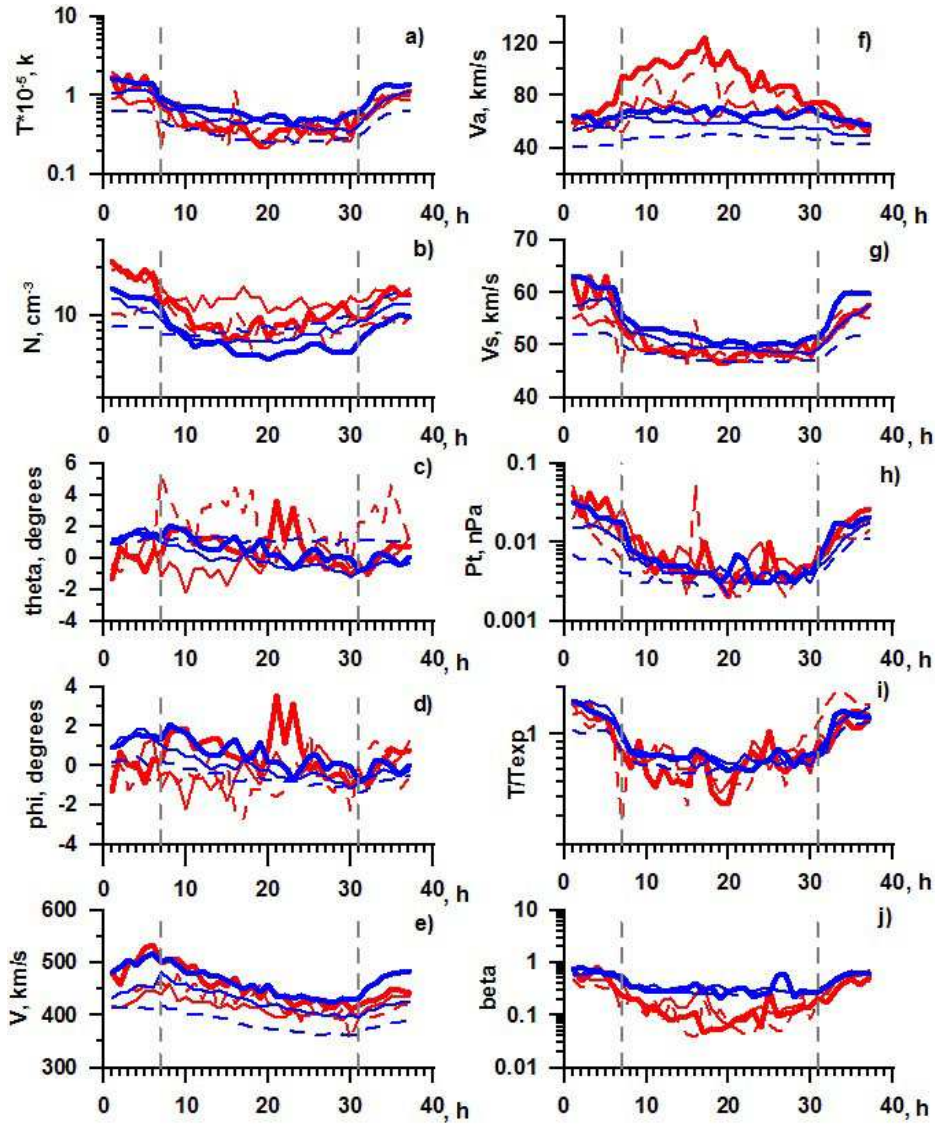}
\caption{The temporal profile of SW plasma and IMF parameters for MC 
(with IS+Sheath, with Sheath and without IS+Sheath - bold, thin and 
dash red lines, respectively), and 
Ejecta (with IS+Sheath, with Sheath and without IS+Sheath - bold, thin 
and dash blue lines) obtained by the double superposed epoch analysis}
\end{figure}

\begin{figure}
\noindent\includegraphics[width=14cm]{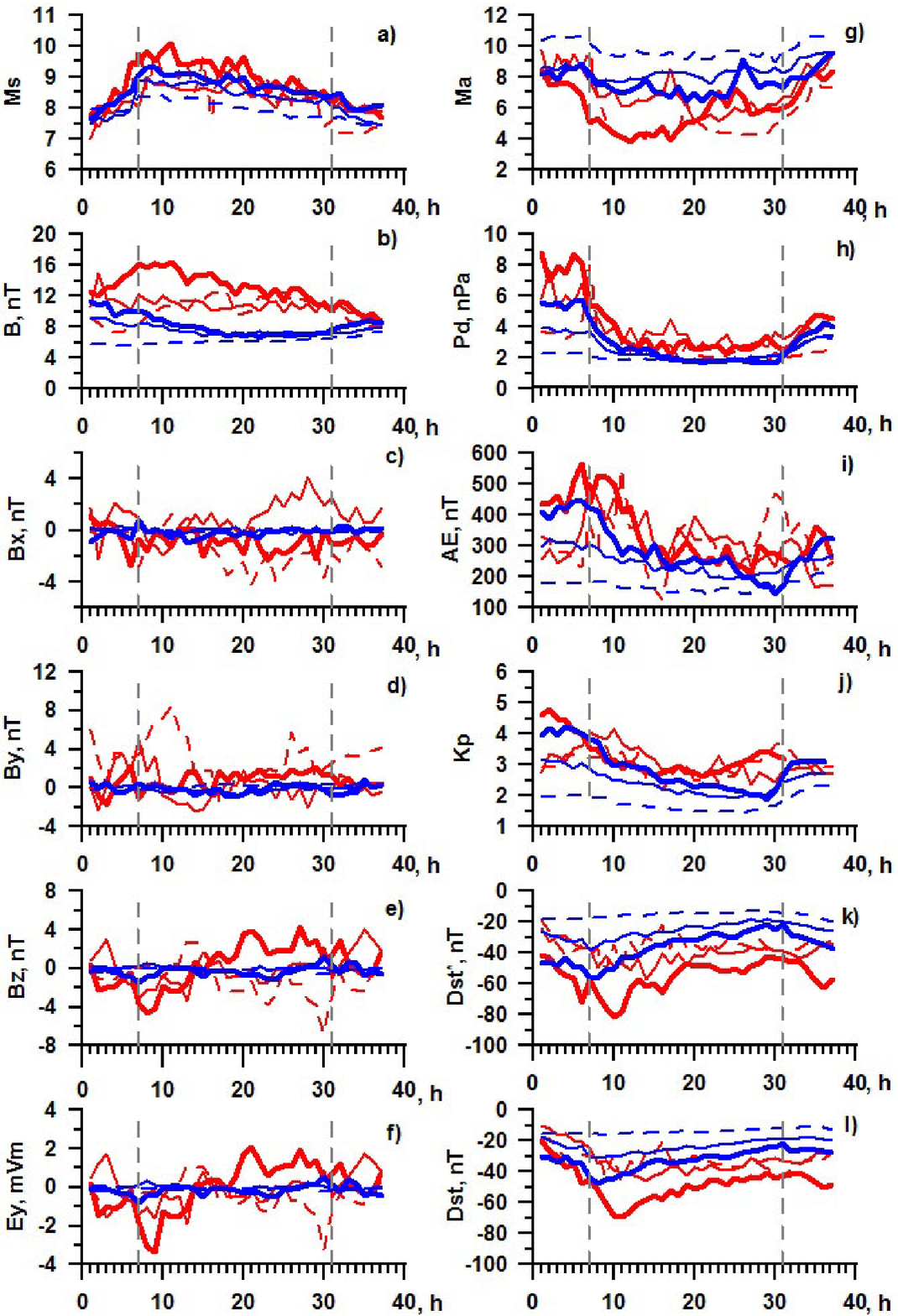}
\caption{Continuation of figure 7}
\end{figure}

%

\end{document}